\documentclass{cim_m}
\newcommand{\ice}[1]{\relax}
\usepackage{graphicx}
\usepackage{epsfig}
\usepackage{latexsym}
\usepackage{amsmath}
\usepackage{amsfonts}
\usepackage{amsxtra}

\newcommand{\slsh}[1]{\not{\hbox{\kern-2pt${#1}$}}}

\newcommand{\ba}[1]{\begin{eqnarray} \label{#1}}
\newcommand{\ea}{\end{eqnarray}}

\def\beq{\begin{equation}}
\def\eeq{\end{equation}}
\def\bea{\begin{eqnarray}}
\def\eea{\end{eqnarray}}
\def\bq{\begin{quote}}
\def\eq{\end{quote}}

\title{$\tau$-Flavour Violation at the LHC} 

\author{M.E.~Gomez\from{a}\ETC,  
E.~Carquin\from{b},
P.~Naranjo\from{a}
\atque
J.~Rodriguez-Quintero\from{a}}
\instlist{
\inst{a} Departamento de F\'{\i}sica Aplicada, University of Huelva, Spain
 \inst{b} Departamento de F\'isica y Centro de Estudios Subat\'omicos,
Universidad T\'ecnica Feder\'ico Santa Mar\'ia, Casilla 110-V, Valpara\'iso, Chi
le
}

\begin{document}

\maketitle
%\begin{quote}
\begin{abstract} 
We study the conditions required for $\chi_2 \to \chi + 
\tau^\pm \mu^\mp$ decays to yield 
observable tau flavour violation at the LHC, for cosmologically interesting values of the neutralino relic density. These condition can be achieved in the framework of a SU(5) model with a see-saw mechanism that allows a possible coexistence of a LHC signal a low prediction for radiative LFV decays.

\end{abstract}

%%%%%%%%%%%%%%%%%%%%%%%%%%%%%%%%%%%%%%%%%%%%
%% MAINMATTER
%%%%%%%%%%%%%%%%%%%%%%%%%%%%%%%%%%%%%%%%%%%%

\section{Introduction}
Data from  both atmospheric~\cite{skatm} and solar~\cite{sksol}
neutrinos have by now confirmed the existence of neutrino oscillations
with near-maximal  $\nu_\mu - \nu_\tau$ mixing (Super-Kamiokande) and
large $\nu_e \to \nu_{\mu}$ one (SNO).
These observations would also imply violation of the
corresponding charged-lepton numbers, which in supersymmetric theories 
might be significant and observable in low-energy
experiments.Many signatures for charged-lepton-flavour 
violation have been considered ~\cite{rev, EGL}, 
including $\mu \to e \gamma$  decays and conversions,
$\tau \to \mu \gamma$ and $\tau \to e
\gamma$ decays.  Other possibilities that have been considered are the decays 
$\chi_2\to \chi + e^\pm \mu^\mp$ \cite{Hisano}, 
and $\chi_2\to \chi + \mu^\pm \tau^\mp$\cite{HP,CEGLR}, 
where $\chi$ is the lightest
neutralino, assumed here to be the lightest supersymmetric particle (LSP),
and $\chi_2$ is the second-lightest neutralino.  We present the results 
from \cite{Carquin}  where we found  that a signal for $\tau$ flavour-violating
$\chi_2$ decays may be observable if the branching ratio exceeds about 10\%.
We consider the cosmologically preferred parameter space (as dictated by WMAP) of $b-\tau$ 
Yukawa-unified models with massive neutrinos  \cite{pedro,mambrini}. 
We find that, assuming general 
structures for the soft terms arising from a horizontal Abelian symmetry, 
$SU(5)$ RGEs efficiently suppress off-diagonal terms in the scalar soft 
matrices \cite{pedro2} as compared 
to the conventional case where the soft terms 
are postulated at the GUT scale, hence rendering the model compatible with 
current experimental bounds.

\begin{figure}
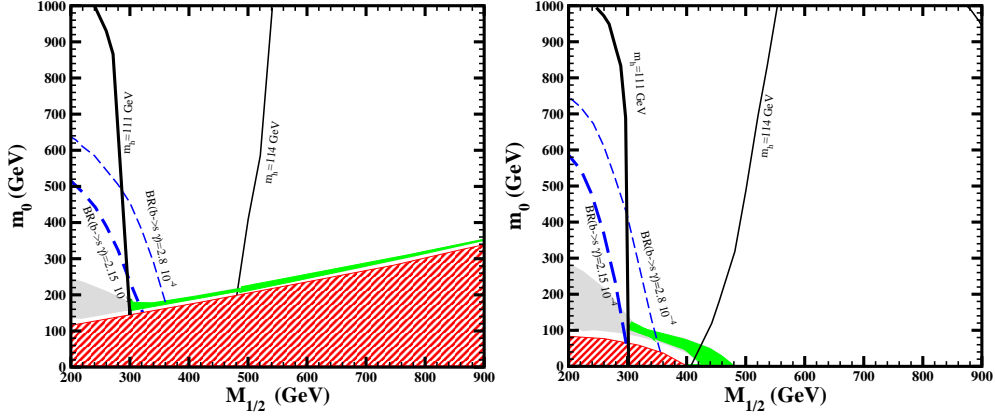

  \includegraphics[width=6.5cm,height=5.5cm]{m0m12_tb35g.eps}
  \includegraphics[width=6.5cm,height=5.5cm ]{m0m12_tb35x.eps} 
\caption{\small \it   Cosmologically-favored areas (green) in 
the $(M_{1/2}, m_0)$ plane for $\tan\beta=35$ and $A_0=m_0$, assuming SU(5) unification. In the left panel we assume universality at $M_X=M_{GUT}$, whereas 
in the right panel we assume universality at $M_X = 2 \times 10^{17}$~GeV. 
The red areas are excluded because $m_\chi> m_{\tilde{\tau}}$. We also display the contours for $m_h=111, 114$~GeV (black solid and thin solid) and 
$BR(b\rightarrow s \gamma)\cdot 10^{4}<2.15, 2.85$ (blue dashed and thin dashed).}
\label{fig:areas}
\end{figure}

\section{Study of SUSY spectrum and parameter space}
We pay particular attention to regions leading
to large values of $\Gamma(\chi_2\rightarrow\chi+ \tau^\pm+ \mu^\mp)$ 
via the on-shell slepton production mechanism:
\begin{equation}
BR(\chi_2\rightarrow\chi \tau^\pm \mu^\mp)=\sum_{i=1}^3 \left[
BR(\chi_2\rightarrow\tilde{l}_i \mu)BR(\tilde{l}_i\rightarrow\tau \chi) + 
BR(\chi_2\rightarrow\tilde{l}_i \tau)BR(\tilde{l}_i\rightarrow\mu \chi)
\right],
\end{equation}
while satisfying all phenomenological
and  cosmological (relic density) constraints. 
The characteristic parameter region for
the signal in the $\tau$ channel to be
optimal is defined by the following:
(i) $m_{\chi_2}> m_{\tilde{\tau}}>m_\chi$;
(ii) one of the mass differences in (i) is $> m_\tau$
and the other $>m_\mu$, $m_{\tilde{\tau}} > m_\chi$, so that the $\mu, \tau$
and $\tilde{\tau}$ are all on-shell;
%(hadronic decays of $\tau$s in the final state);
(iii) moderate values of $m_\chi$ (phase space and luminosity considerations).

Fig. \ref{fig:areas} is used to select points satisfying all phenomenological 
constraints for the event analysis of the next section. In particular,  
for $SU(5)$ unification and assuming 
universal soft terms at $M_X=2\times 10^{17}$ GeV, $\tan\beta=35$ is the 
smallest value of $\tan\beta$ such that the WMAP area in the plane 
$M_{1/2}-m_0$ is not excluded by the $m_h$ bound \cite{mambrini, pedro}. 

In Table~1 we display parameters of the two reference points A and B. 
Point A is the CMSSM model used in \cite{HP}. Point B is a model with universality assumed at a scale $2\cdot 10^{17}$~GeV;  
for comparison with this point, we also present point C,
a set of CMSSM parameters that leads to a similar sparticle
spectrum and satisfies all the cosmological
and phenomenological bounds. In all cases, we work with $\mu>0$.

\begin{table}[!h]
%\begin{center}
\begin{tabular}{| c | c | c | c | c | c | c | c | c |}
\hline
$Point$ & $Model type$ & $m_0$ & $M_{1/2}$ & $\tan\beta$ & $A_0$ & $N_{events}$ & $\sigma_{int}$ & $L_{int}$ \\
\hline
%\hline
A & CMSSM & $100$ & $300$ & $10$ & $300$ & 757K & 25.3~pb & 30~fb$^{-1}$ \\
\hline
B & SU(5) & $40$ & $450$ & $35$ & $40$ & 730 K & 2.44~pb & 300~fb$^{-1}$\\ 
\hline
C & CMSSM & $220$ & $500$ & $35$ & $220$ & 536 K & 1.79~pb & 300~fb$^{-1}$ \\
\hline
\end{tabular}
\caption{}
\label{tab:a}
\end{table}

\section{SUSY Lepton Flavour Violation}
The flavour mixing entries are defined as:
\beq
\delta_{XX}^{ij}=(M^2_{XX})^{ij}/(M^2_{XX})^{ii}\;\;\;\;\;\;(X=L,R).
\eeq
We take into account only $2-3$ generation 
flavor mixing. 
%The evaluation of the LFV observables is done by performing
%a full diagonalization of the slepton mass matrices.

\begin{figure}[!t]
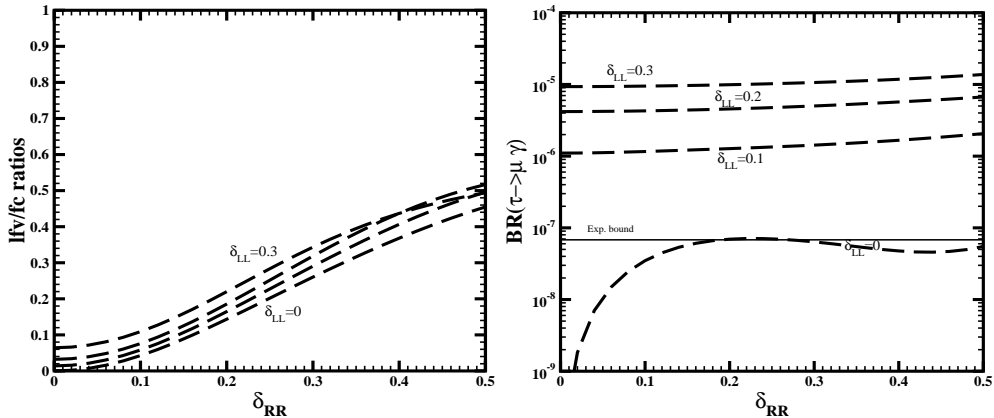

  \includegraphics[width=6.5cm,height=5.5cm]{ratdel_RRB.eps}
  \includegraphics[width=6.5cm,height=5.5cm]{brdel_RRB.eps} 
\caption{\small \it  Branching ratios for point B as functions of 
$\delta_{RR}$ for certain discrete choices of $\delta_{LL}$}.
\label{ratdel}
\end{figure}

In order to have significant LFV signals, we need 
$\Gamma(\chi_2\rightarrow\chi+ \tau^\pm+ \mu^\mp)/
\Gamma(\chi_2\rightarrow\chi+ \tau^\pm+ \tau^\mp) \sim 0.1$. Values of 
$\delta_{LL}$ leading to these ratios would imply a significant violation of the$\tau\rightarrow \mu \gamma$ bound.  In Fig.~\ref{ratdel} we present 
the dependence of these decays with  the flavour mixing parameters 
$\delta_{RR}$ and $\delta_{LL}$ for point B. We see that in this 
case we need large non-diagonal entries in the slepton mass matrix in order 
to achieve a
branching ratio for $\tilde{\chi}_2\rightarrow \tilde{\chi}_1\tau^\pm\mu^\mp$ that is
of interest for the LHC, e.g., $\delta_{RR} \sim 0.15$ for $\delta_{LL} = 0$
or $\delta_{LL} \sim 0.35$ for $\delta_{RR} = 0$. We also see 
that $\tau\rightarrow \mu \gamma$ is very restrictive on the size of 
$\delta_{LL}$, imposing a maximum value $\sim 0.03$. We see in the 
bottom-right
panel that $\delta_{RR} \sim 0.15$ is allowed for $\delta_{LL} = 0$. Due to the strong bound imposed by $\tau\rightarrow \mu \gamma$, 
it is very difficult to obtain reasonable values 
of  $\delta$ using  only the LL  and/or RL mixing found in seesaw models. However, significant FV entries on the RR sector can 
be generated only by using non-minimal models for the soft terms. 

\begin{figure}[!h]
\begin{center}
\hspace*{-0.3 cm}
\includegraphics[width=10cm,height=8cm]{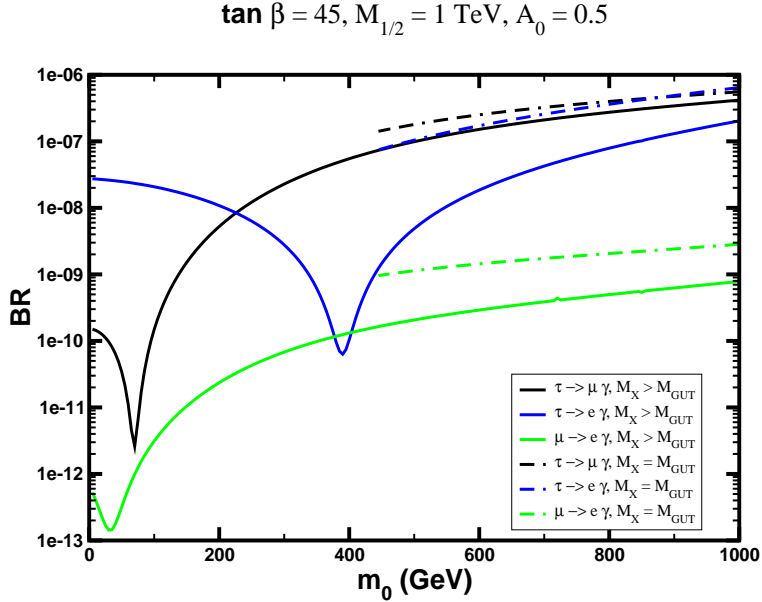}
\caption{ \it Prediction for the charged-lepton flavour violating branching 
ratios showing the difference of taking either $M_X$ or $M_{GUT}$ as the high 
scale. 
}
\label{MXMGUT}
\end{center}
\end{figure}

\section{SU(5) Unification and GUT soft masses }

The introduction of non-trivial flavour structures for the slepton soft 
terms at $M_{GUT}$, although being reasonable as an implication of the 
family symmetry responsible for the Yukawa texture, typically results on a large violation of the bounds on $\l_j\rightarrow l_i \gamma$ \cite{Leontaris, Chankowski}. This picture may be remedied if we assume that SUSY is broken with universal soft terms at a scale 
$M_X > M_{GUT}$. In this case, the cosmological requirement of 
having a neutral particle as the LSP imposes low values on $m_0$, such that 
$m_{\tilde{\tau}}>m_\chi$ \cite{pedro,mambrini}, since diagonal terms of the soft masses have a large RGE growth, while  non-diagonal elements remain almost unaffected by the runs. Thus, even assuming non-diagonal soft terms 
with matrix elements of the same order of magnitude at $M_X$, the corresponding matrix at $M_{GUT}$ exhibits dominant diagonal elements. To some extent, the RGE effect is similar to the action of closing an umbrella: the general non-
universal soft terms at $M_X$ resemble an open umbrella that approaches a diagonal matrix at the GUT scale.

In Fig.~\ref{MXMGUT}, we show the differences respect considering: 
i) SU(5) RGE evolution of the soft terms from the high scale $M_X$ down to $M_{GUT}$ and then the MSSM with see-saw neutrinos (solid lines),  ii) Soft SUSY breaking terms given 
at $M_{GUT}$ and then the MSSM  with see-saw neutrinos (dash-lines).
In case ii) we stop the lines at the value of $m_0$ below which $m_{\tilde{\tau}}$ becomes the LSP. In contrast, $m_0$ can even vanish at $M_X$ in case i). 
We used the same textures and soft terms as in Ref.~\cite{Chankowski}. However, unlike these authors,  we decouple the right-handed neutrinos 
below $M_{GUT}$. As a result, the predicted BR's do not vanish in the limit 
$m_0=0$. 

We can provide one explicit example of the growth of the diagonal terms 
of the slepton mass matrix in models with interesting predictions for both 
LFV and $\Omega_\chi h^2$.  Let us consider the $0<m_0<100$~GeV region. 
In the area of the parameter space where WMAP bounds are 
satisfied due to $\tau-\chi$ coannihilations, we find that 
$m_{1/2}$ obeys a linear function of $m_0$, $m_{1/2}\sim a_1^i+a_2^im_0$, 
where $i$ runs over the multiplets. It turns out that, taking into account 
that the radiative corrections to the off-diagonal entries of the soft mass 
matrices are subdominant as compared with those of the diagonal ones, these 
diagonal elements can be expressed as follows:
\begin{equation}
m_{S_i}^2\simeq C_i^2\left(m_0\right)m_0^2,
\label{umb}
\end{equation}
where we have defined
\begin{equation}
C_i^2\left(m_0\right)\equiv \frac{144}{20\pi}\alpha _5\left(\left(
\frac{a_1^i}{m_0}\right)^2+\frac{2a_1^ia_2^i}{m_0}+a_2^2\right)\ln\left(
\frac{M_X}{M_{GUT}}\right)
\end{equation}
and $S_i$ stands for the supermultiplets $\mathbf{10}$ and $\mathbf{\bar{5}}$. 
As stated, Eq.(\ref{umb}) implies a large enhancement only of the diagonal 
entries of the soft matrices, thus further suppressing the off-diagonal 
elements. It turns out indeed that for values of $m_0\simeq 60-80$ GeV at 
$M_X$ such an enhancement at the GUT scale is as large as $\simeq$ 100. As a 
consequence, the soft mass matrices $\bar{m}_{10}^2$ and 
$\bar{m}_5^2$ at GUT scale read as
\begin{equation}
\bar{m}_{10}^2=\left(\begin{array}{ccc}
1 & \varepsilon ^3 & \varepsilon ^5 \\
\varepsilon ^3 & 1 & \varepsilon ^4 \\
\varepsilon ^5 & \varepsilon ^4 & 1 \\
\end{array}\right)C^2\left(m_0\right)m_0^2,\,\,\,\,\,
\bar{m}_5^2=\left(\begin{array}{ccc}
0 & 0 & 0 \\
0 & 1 & \varepsilon ^3 \\
0 & \varepsilon ^3 & 1 \\
\end{array}\right)C^2\left(m_0\right)m_0^2
\label{matrix}
\end{equation}

\begin{figure}[!h]
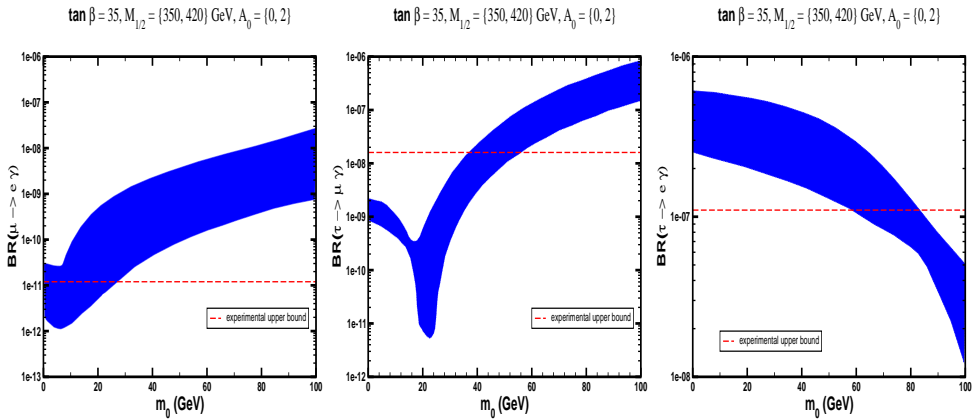

\begin{center}
\hspace*{-0.3 cm}
\includegraphics[width=4.2cm,height=5.5cm]{BR_meg_35_1_final.eps}
\includegraphics[width=4.2cm,height=5.5cm]{BR_tmg_35_1_final.eps}
\includegraphics[width=4.2cm,height=5.5cm]{BR_teg_35_1_final.eps}
\caption{ \it Prediction for the $l_j\rightarrow l_i\gamma$  branching ratios 
for the cosmologically preferred area of values of $M_{1/2},\,A_0$, for choice of $U(1)_F$ charges in Ref.~\cite{pedro2} leading to the matrices of Eq.~\ref{matrix}.
}
\label{LFV35}
\end{center}
\end{figure}

which clearly exhibit the suppression on the off-diagonal terms (compare 
with textures. The corresponding predictions for LFV radiative decays at $\tan\beta=35$ are displayed in Fig.\ref{LFV35}.

\section{SUSY-LFV events at the LHC}
\begin{figure}
  \includegraphics[height=.33\textheight]{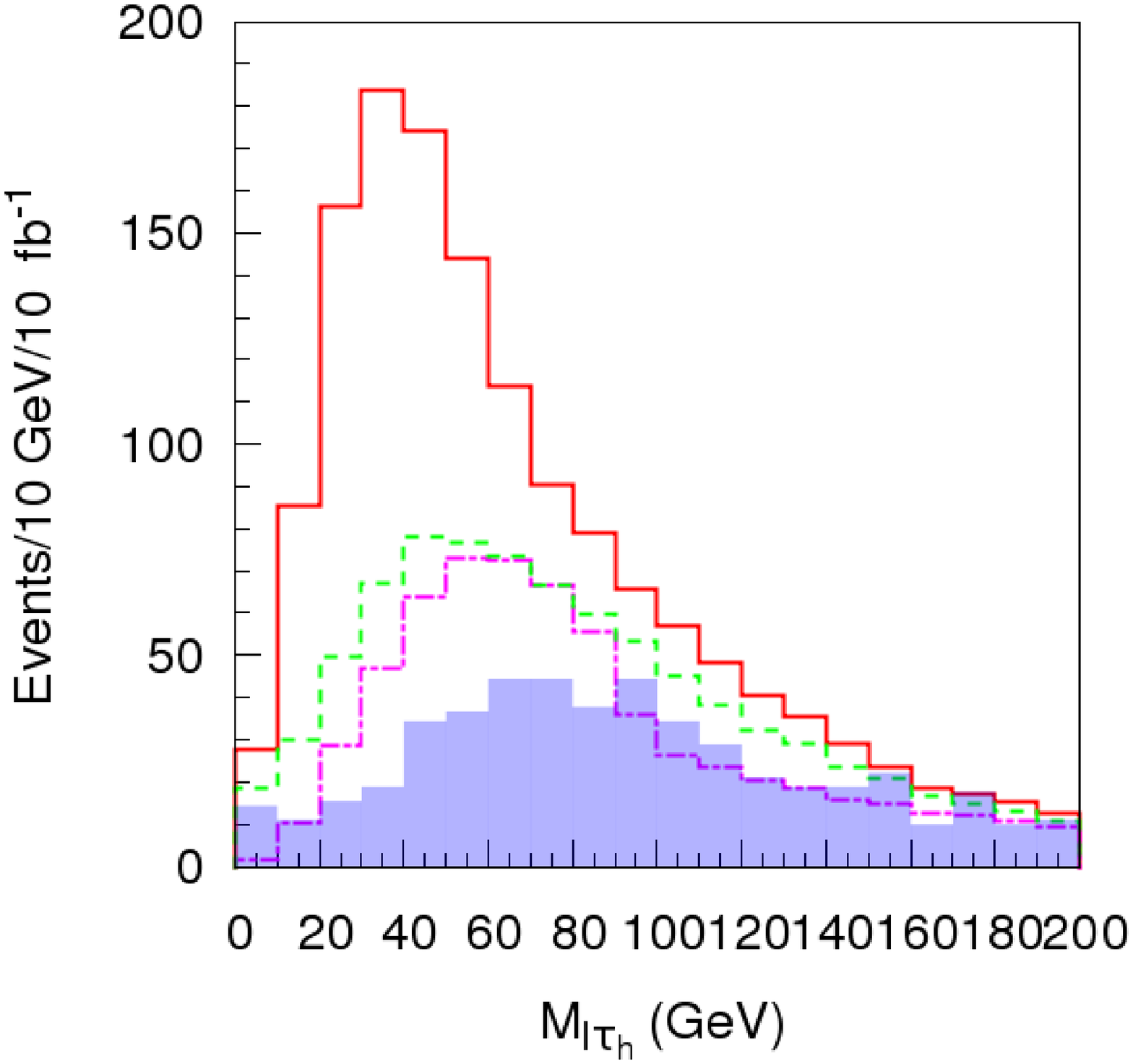}
  \includegraphics[height=.33\textheight]{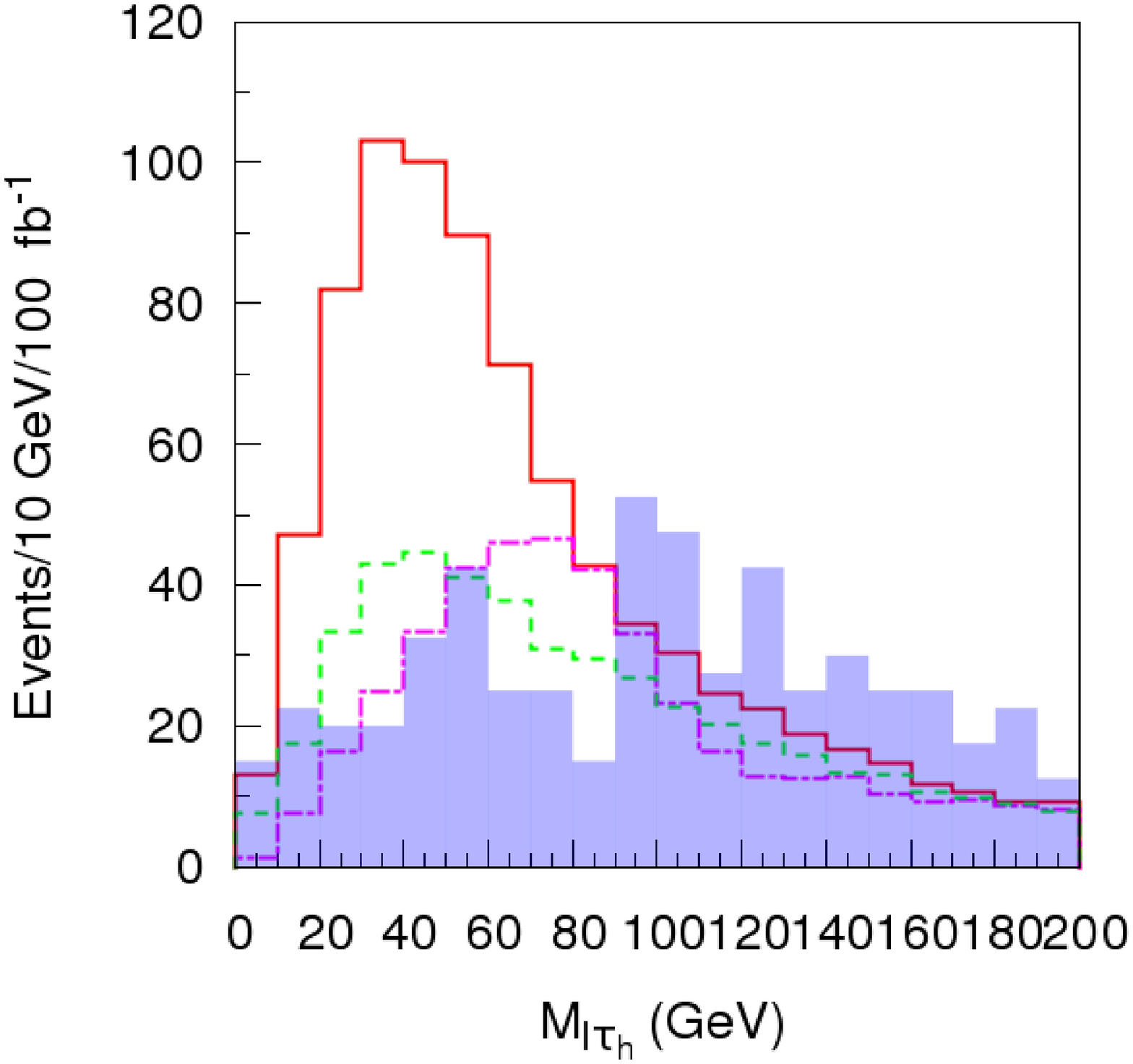}
\caption{\small \it  Visible $l\tau_h$ mass distributions with OS 
(red solid lines), 
SS (green dashed lines), and OS Standard Model backgrounds (shaded).
The LFV OS $\mu\tau_h$ pairs are also shown (pink-dot-dashed). Left plot is for case A and right plot for case B.} 
\label{fig:h4}
\end{figure}

The spectra for points of Table~1 were calculated 
using {\tt ISAJET~7.78}~\cite{ISAJET} and then interfaced into {\tt PYTHIA~6.418}~\cite{Pytia}. The SUSY-LFV decays described in the 
previous sections give a $\mu^\pm \tau^\mp$ pair
and produce an asymmetry between  $\mu^\pm \tau^\mp$ and $e^\pm \tau^\mp$
final states that would not be observable in the case of 
charged lepton number conservation. In Fig.~\ref{fig:h4}, 
an excess of OS $l^{\mp}\tau_h^{\pm}$ pairs over the SS pairs can be seen. 
The left (right) plot  corresponds to point A (B) and shows 
the numbers of events normalized
to a reference luminosity of 10~fb$^{-1}$ (100~fb$^{-1}$). The observable numbers, $N_{\mu\tau_h}^{lfv}$, of $\mu^{\mp}\tau^{\pm}_h$ LFV pairs are obtained 
by summing the counts in the subtracted
$\mu^{\mp}\tau^{\pm}_h - e^{\mp}\tau^{\pm}_h$ distributions in
the interval of $M_{l\tau}$ masses between $30$ and $110$~GeV. We obtain
\begin{eqnarray}
{\rm Point~A}: \; N_{\mu\tau_h}^{lfv}&=&470 \ \pm 39 \ (12 \ \sigma)
\nonumber \\
 {\rm Point~B}: \; N_{\mu\tau_h}^{lfv}&=&308 \ \pm 30 \ (10 \ \sigma)
\end{eqnarray}
where we quote only the statistical errors for the signal samples. 
If we estimate an efficiency of
70 \% for the jet-tau matching, the signal is reduced to $10\sigma$ for 
point A and  $9\sigma$ for point B. Therefore, LFV signal has a good likelihood of being observable, as
long as its branching ratio exceeds about 10\%.

\section{Conclusions}
The observation of LFV in neutralino decays at the LHC can 
be possible if ${\Gamma(\chi_{2}\rightarrow\chi_{1}\tau^{\pm}\mu^{\mp})}$
$/{\Gamma(\chi_{2}\rightarrow\chi_{1}\tau^{\pm}\tau^{\mp})}\sim 0.1$. The LFV 
signal remains observable at points which are favoured in the usual 
CMSSM framework. Finally, we conclude that the search for this 
decay at the LHC is interesting and complementary to the parallel searches for
$\tau \to \mu \gamma$ decays, for non-minimal GUTs.  Furthermore, linear 
colliders will allow to explore complementary parameter space \cite{carquin2}. 
  
As a final remark, let us stress that the phenomenological analysis performed 
in this work can naturally be embedded within a $SU(5)$ GUT model featuring 
non-universal soft terms at the high scale $M_X$, whose origin can be traced 
back to a $U(1)_F$ family symmetry \cite{pedro2}.

%%%%%%%%%%%%%%%%%%%%%%%%%%%%%%%%%%%%%%%%%%%%%%%%
%% BACKMATTER
%%%%%%%%%%%%%%%%%%%%%%%%%%%%%%%%%%%%%%%%%%%%%%%%

%\begin{theacknowledgments}
{\bf Acknowledgments}

The work of E. C. has been partly supported by the MECESUP Chile Grant and the HELEN Program. M.E.G and J.R.Q are supported by the
Spanish MICINN projects FPA2009-10773, FPA2008--04063--E/INFN and the project P07FQM02962 funded
by ``Junta de Andalucia''. 

%\end{theacknowledgments}



\bibliographystyle{aipprocl} % if natbib is missing

%\bibliography{sample}

\begin{thebibliography}{9}

\bibitem{skatm}\BY{
Fukuda~Y. {\it et al.} [Super-Kamiokande Collaboration]} 
\IN{Phys. Rev. Lett.}{81}{1998}{1562}

\bibitem{sksol}\BY{
Fukuda~Y. {\it et al.} [Super-Kamiokande Collaboration]}
\IN{Phys. Rev. Lett.}{82}{1999}{1810};
\IN{Phys. Rev. Lett.}{82}{1999}{2430};
\BY{
Ahmad~Q.~R. {\it et al.} [SNO Collaboration]}
\IN{Phys. Rev. Lett.}{87}{2001}{071301}

\bibitem{rev}\BY{
Kuno~Y. \atque  Okada~Y.} 
\IN{Rev. Mod. Phys.}{73}{2001}{151}

\bibitem{EGL}\BY{
Ellis~J., Gomez~M.E. \atque Lola~S.}
\IN{JHEP}{0707}{2007}{052}


\bibitem{Hisano}\BY{
Hisano~J., Nojiri~M.M., Shimizu~Y. \atque Tanaka~M.}
\IN{Phys. Rev. D}{60}{1999}{055008}

\bibitem{HP}\BY{
Hinchliffe~I. \atque Paige~F.E.}
\IN{Phys. Rev. D}{63}{2001}{115006}

\bibitem{CEGLR}\BY{
Carvalho~D., Ellis~J., Gomez~M., Lola~S. \atque Romao~J.}
\IN{Phys. Lett. B}{618}{2005}{162}


\bibitem{Carquin}\BY{
Carquin~E., Ellis~J., Gomez~M.E., Lola~S. \atque Rodriguez-Quintero~J.}
  %``Search for Tau Flavour Violation at the LHC,''
\IN{JHEP}{0905}{2009}{026}
%  [arXiv:0812.4243 [hep-ph]].
  %%CITATION = JHEPA,0905,026;%%


\bibitem{pedro}\BY{
Gomez~M.E., Lola~S., Naranjo~P. \atque Rodriguez-Quintero~J.}
  %``WMAP Dark Matter Constraints on Yukawa Unification with Massive
  %Neutrinos,''
\IN{JHEP}{0904}{2009}{043}
%  [arXiv:0901.4013 [hep-ph]].
  %%CITATION = JHEPA,0904,043;%%


\bibitem{mambrini}\BY{
Calibbi~L., Mambrini~Y. \atque Vempati~S.K.}
  %``SUSY-GUTs, SUSY-Seesaw and the Neutralino Dark Matter,''
\IN{JHEP}{0709}{2007}{081}
 % [arXiv:0704.3518 [hep-ph]].
  %%CITATION = JHEPA,0709,081.

\bibitem{pedro2}\BY{
Gomez~M.E., Lola~S., Naranjo~P. \atque Rodriguez-Quintero~J.}
. %``Suppression of Lepton Flavour Violation from Quantum Corrections above
  %$M_{GUT}$,''
  arXiv:1003.4937 [hep-ph].
  %%CITATION = ARXIV:1003.4937;%%

\bibitem{Leontaris}\BY{
Leontaris~G.K. \atque Tracas~N.D.}
  %``Lepton flavour violation in unified models with U(1)-family symmetries,''
\IN{Phys. Lett. B}{431}{1998}{90};
%  [arXiv:hep-ph/9803320], 
  %%CITATION = PHLTA,B431,90;%%
\BY{
Gomez~M.E., Leontaris~G.K., Lola~S. \atque Vergados~J.D.}
  %``U(1)-textures and lepton flavor violation,''
\IN{Phys. Rev. D}{59}{1999}{116009}
%  [arXiv:hep-ph/9810291].
  %%CITATION = PHRVA,D59,116009;%%

\bibitem{Chankowski}\BY{
Chankowski~P.H., Kowalska~K., Lavignac~S. \atque Pokorski~S.}
arXiv:hep-ph/0507133 


\bibitem{ISAJET}\BY{
Baer~H., Paige~F.E., Protopopescu~S.D. \atque Tata~X.}
arXiv:hep-ph/0312045

\bibitem{Pytia}\BY{
Sjostrand~T., Mrenna~S. \atque Skands~P.}
\IN{JHEP}{05}{2006}{026}
(LU TP 06-13, FERMILAB-PUB-06-052-CD-T)

\bibitem{carquin2}\BY{
Carquin~E., Ellis~J., Gomez~M.E., Lola~S. \atque Rodriguez-Quintero~J.}
  %``Search for Tau Flavour Violation at the LC,''
(in preparation).




\end{thebibliography}

%  \typeout{******************************************}
%  \typeout{** Please run "bibtex \jobname" to optain}
%  \typeout{** the bibliography and then re-run LaTeX}
%  \typeout{** twice to fix the references!}
%  \typeout{******************************************}

%\end{document}

%%%%%%%%%%%%%%%%%%%%%%%%%%%%%%%%%%%%%%%%%%%
%% The following lines show an example how to produce a bibliography
%% without the help of the BibTeX program. This could be used instead
%% of the above.
%%%%%%%%%%%%%%%%%%%%%%%%%%%%%%%%%%%%%%%%%%%

\end{document}